# Active shape correction of a thin glass/plastic X-ray mirror

D. Spiga[1§], M. Barbera[2], S. Basso[1], M. Civitani[1], A. Collura[3], S. Dell'Agostino[1], U. Lo Cicero[3], G. Lullo[2], C. Pelliciari[1], M. Riva[1], B. Salmaso[1], L. Sciortino[2]

[1]INAF - Brera Astronomical Observatory, Via Bianchi 46, 23807 Merate  (Italy)
[2]Università degli Studi di Palermo, Via Archirafi 36, 90123 Palermo (Italy)
[3]INAF - Osservatorio Astronomico di Palermo, Piazza del Parlamento 1, 90134 Palermo (Italy)

## ABSTRACT

Optics for future X-ray telescopes will be characterized by very large aperture and focal length, and will be made of lightweight materials like glass or plastic in order to keep the total mass within acceptable limits. Optics based on thin slumped glass foils are currently in use in the NuSTAR telescope and are being developed at various institutes like INAF/OAB, aiming at improving the angular resolution to a few arcsec HEW. Another possibility would be the use of thin plastic foils, being developed at SAO and the Palermo University. Even if relevant progresses in the achieved angular resolution were recently made, a viable possibility to further improve the mirror figure would be the application of piezoelectric actuators onto the non-optical side of the mirrors. In fact, thin mirrors are prone to deform, so they require a careful integration to avoid deformations and even correct forming errors. This however offers the possibility to actively correct the residual deformation. Even if other groups are already at work on this idea, we are pursuing the concept of active integration of thin glass or plastic foils with piezoelectric patches, fed by voltages driven by the feedback provided by X-rays, in intra-focal setup at the XACT facility at INAF/OAPA. In this work, we show the preliminary simulations and the first steps taken in this project.

## 1. INTRODUCTION

X-ray optics of the future will need to conjugate very large effective areas and high angular resolutions, for example ATHENA, selected for the L2 slot in ESA's Cosmic Vision 2015–25 with a launch foreseen in 2028, requests an effective area of 2 m$^2$ and a HEW (*Half Energy Width*) < 5 arcsec at 1 keV. The available techniques for X-ray astronomical mirror manufacturing have so far privileged either requirement. Thick and stiff mirrors with high accuracy figuring and polishing, like Chandra's, enable an excellent angular resolution (HEW $\approx$ 0.5 arcsec) but only a low filling of the available aperture, since only a small number of mirrors can be assembled. This obviously goes at the expense of the effective area that can be reached with a given mass of the optics. A denser mirror nesting, as in the case of XMM-Newton or Swift-XRT, requires the adoption of thinner mirrors (with thickness on the order of 1 mm or less). This clearly yields higher effective areas, but makes the mirrors more flexible and prone to deform, and the angular resolution degrades in proportion. Nowadays, wide-aperture optical modules cannot be made of monolithic mirrors, but have to be based on the assembly of modular elements (XOU, *X-ray Optical Units*) obtained stacking thin mirrors. In addition, keeping the mass to within acceptable limits for a few meters diameter optics requires lightweight materials like Silicon, glass or plastic thin foils.

The development of lightweight modular optics for ATHENA (formerly IXO and XEUS) at ESA/ESTEC is ongoing since 2004, based on the Silicon Pore Optics technology[1]. However, a backup technology for IXO/ATHENA optics has been developed at INAF/OAB under ESA/ESTEC contract in 2009-2013 and in parallel at MPE, based on the hot slumping of thin glass foils. This approach was already used to build the optics of the NuSTAR hard X-ray telescope[2], currently in operation. Owing to the low density of both silicon and glass, and to the small thickness of the foils at play, both technologies are suitable to provide the required effective area/mass ratio, and the efforts are concentrated on the improvement of the imaging quality. For slumped glass foils[3], this action is still ongoing on two main fronts: minimization of the surface roughness and minimization of profile errors. The former is obtained starting from highly polished glass substrates, most of them commercially available, and avoiding the surface degradation stemming from contact with the moulds used to impart the correct shape to the mirrors, or, if a contact is needed, minimizing the shear of the optical surface on the mould. The latter can be improved by a proper optimization of the parameters used in the thermal process[3], adopting precisely figured slumping and integration moulds, and along with a dedicated integration process[5], aimed at suppressing most of the long-period errors remaining in the glass foil profiles after the slumping process. Other deformations may arise post integration as a result from mechanical stresses caused by the accelerations experienced at launch, or as a result of a thermal load applied to a small, but nonzero, CTE mismatch between the materials composing the XOU[6].

---

[§] contact author: daniele.spiga@brera.inaf.it

A research that has been carried out in parallel for the development of lightweight focusing mirrors is based on the technology of plastic foils[7],[8],[9], carried out by a collaboration between SAO, Palermo University, INAF/OAPA, DSRI. Also in this case the plastic material is flexible and lightweight, so it can be easily formed to build a focusing shape. Moreover, it has been proven to reach good roughness levels also after the deposition of a reflective coating[10]. Also in this case, the problem to be faced is the shape accuracy that can be achieved constraining the plastic foil to a finite number of locations.

Even if a constant progress has been made in those years, the angular resolutions achieved are not yet at the 5 arcsec HEW required for ATHENA, and even farther from the sub-arcsec resolutions envisaged by some mission concepts[11]. Several improvements may be expected from the advancing material technology (as indicated in the recent years by the newly developed kinds of glasses with superb mechanical properties, e.g., Gorilla® and Willow® glasses by Corning). An alternative solution may be the adoption of a system of piezoelectric actuators applied to the non-optical side of the thin foils, in order to actively correct the shape at the needed locations. Other groups are already studying this approach, with results extensively described in some other papers of this volume[12],[13],[14]. However, the existing adjustable optical systems rely on a feedback based on metrology tools that are difficult to use in a densely stacked XOU. The concept we propose here is based on a matrix of commercially available actuators applied on the backside of a glass or plastic mirror, but with the piezo array voltage determined *in situ*, reconstructing the mirror profile in X-rays by intra-focal imaging[15]. This method does not require us to inspect the optical surface with metrological tools and can in principle be utilized also in operational conditions, i.e., using an astronomical X-ray source to optimize the mirror shape, and for consequence the angular resolution.

In this paper we describe the initial activities performed at INAF/OAB, Università di Palermo, and INAF/OAPA, in the context of the AXYOM (*Adjustable X-raY optics for astrOnoMy*) project, aimed at improving the shapes of mirrors in slumped glass and plastic foils developed in the last years, endowing them with a piezoelectric actuation. In Sect. 2 we recapitulate the results achieved in the last years through the development of static optics based on glass and plastic foils. We describe the activities being initiated in the AXYOM project in Sect. 3. In Sect. 4 we describe the intra-focal imaging method we will use at the XACT[16],[17] facility (INAF/OAPA) to characterize the optical prototypes in X-rays and to give the feedback to the control electronics. Finally, Sect. 5 drafts some preliminary conclusions.

## 2. GLASS/PLASTIC THIN FOIL MIRRORS

### 2.1. Thin glass foil optics

The hot slumping forming technique for thin glass foils has been adopted at INAF/OAB to figure lightweight X-ray mirrors as a backup technology for the IXO telescope. It consists of a direct slumping replication of a precisely figured cylindrical mould, with the concave optical side in contact with the mould surface, in a similar way to the mirrors fabricated for NuSTAR[2]. An advantage is that the final profile is largely independent of the glass thickness variations. In contrast, the surface roughness can be damaged by the contact with the mould, and the profile is altered if dust is trapped between the glass and the mould. The advantages are reversed if the indirect process is adopted (e.g., at MPE[18]).

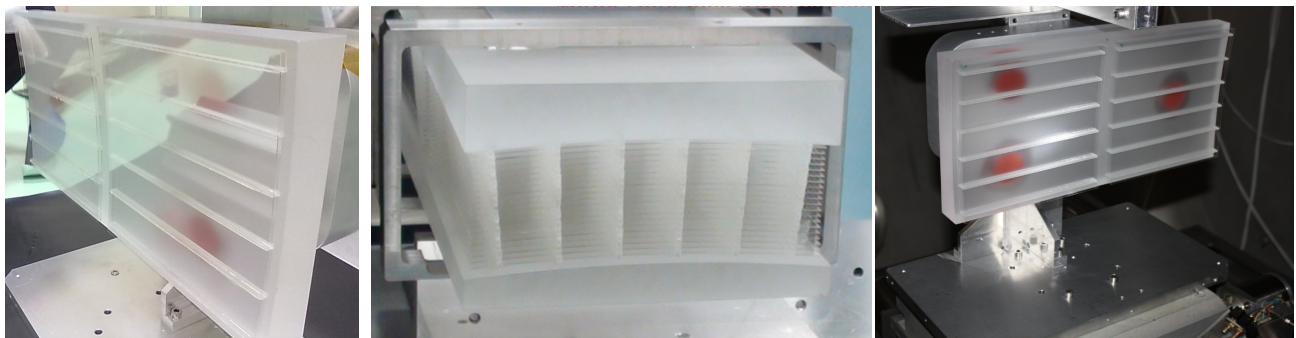

Fig. 1: Some X-ray optic prototypes developed at INAF/OAB based on hot slumped glasses (after[5]) under ESA/ESTEC contract. (left) PoC#1 (*Proof of Concept* No. 1), 2 plate pairs. (center) XOU_BB, 20 plate pairs. (right) PoC#2, 4 plate pairs.

Table 1. The demonstrators manufactured for the "IXO backup optics with slumped glasses" project[15], with the X-ray tests performed at PANTER at the 0.27 keV X-ray energy, with the methods used to assess the HEW.

| Item | date | tested | single shell result | method |
|---|---|---|---|---|
| *PoC#1* | Dec 2011 | intrafocus | HEW ≈ 80 arcsec | from surface metrology |
| *XOU_BB* | Aug 2012 | in focus | HEW ≈ 60 arcsec | measured in X-rays |
| *PoC#2* | Apr 2013 | intrafocus | HEW = 17 arcsec | reconstructed in X-rays |
| *PoC#2* | Feb 2014 | in focus | HEW ≈ 20 arcsec | measured in X-rays |

In addition to the hot slumping process adopted for NuSTAR, simply driven by gravity, the INAF/OAB process makes use of pressure[3] exerted on the glass foil for a better replication of the slumping mould (INAF patent). Moreover, the glass (AF32® by Schott or EAGLE XG® by Corning) and the mould (K20 ceramic) are chosen with the best possible CTE matching in order to minimize the shear between the optical surface of the glass and the mould. In this way, the surface roughness degradation is limited to tolerable values up to the energy of 1 keV. Pairs of cylindrical foils are subsequently integrated onto precisely figured parabolic / hyperbolic moulds in Fused Silica to reconstruct the Wolter-I profile[19], widespread in X-ray astronomical mirrors. To this end, the cylindrical shape is forced against the integration mould and the imparted Wolter-I shape is frozen by gluing stiffening ribs on the non-optical side, previously fixed to a still backplane or to another mirror of the stack[20]. After the separation from the mould, the glass maintains the parabola-hyperbola profile in correspondence to the ribs, while it tends to return to its original shape in the space between the ribs, resulting in a figure error with respect to the exact Wolter-I shape. Nevertheless, for the prototypes produced with a 20 m focal length, 1 m curvature radius, and 0.4 mm thickness, the spring-back effect amounts to less than 0.5 arcsec if the slumped glass is a perfect cylindrical sector[21]. Another couple of cylindrical foils can be integrated onto the previous mirror, building a stacked XOU. In Fig. 1 we show some demonstrators produced at INAF/OAB as part of this activity.

However, since slumping defects are always present, the deviation from a perfect cylinder will be partly retained after integration. According to Finite Element Modeling (FEM), the fraction of error amplitude that is *not* corrected by the integration process depends on the distance from the ribs and the spatial frequency of the error in the longitudinal direction (azimuthal errors have a lesser weight by two orders of magnitude and are negligible in this case). It takes on a maximum amid two consecutive ribs, and it increases with the spatial frequency: low frequency errors are mostly suppressed, while errors in the centimeter range remain essentially unchanged. The result is a profile error that, together with the surface roughness, entirely determines the mirror HEW. In Tab. 1 we list the angular resolution performances, predicted or directly measured in X-rays at the PANTER facility[22] (MPE, Germany), for the demonstrators realized. The profile error can be measured using a surface-mapping tool and exhibits the typical behavior shown in Fig. 2, left. The spring-back effect, maximum in between the ribs along the longitudinal direction, is the dominant source of errors. Also visible are defects in the centimeter range of spatial wavelengths. An analogous picture (Fig. 2, right) comes from the intra-focal image of the same mirror measured at PANTER (Fig. 7), following the method described in Sect. 4.

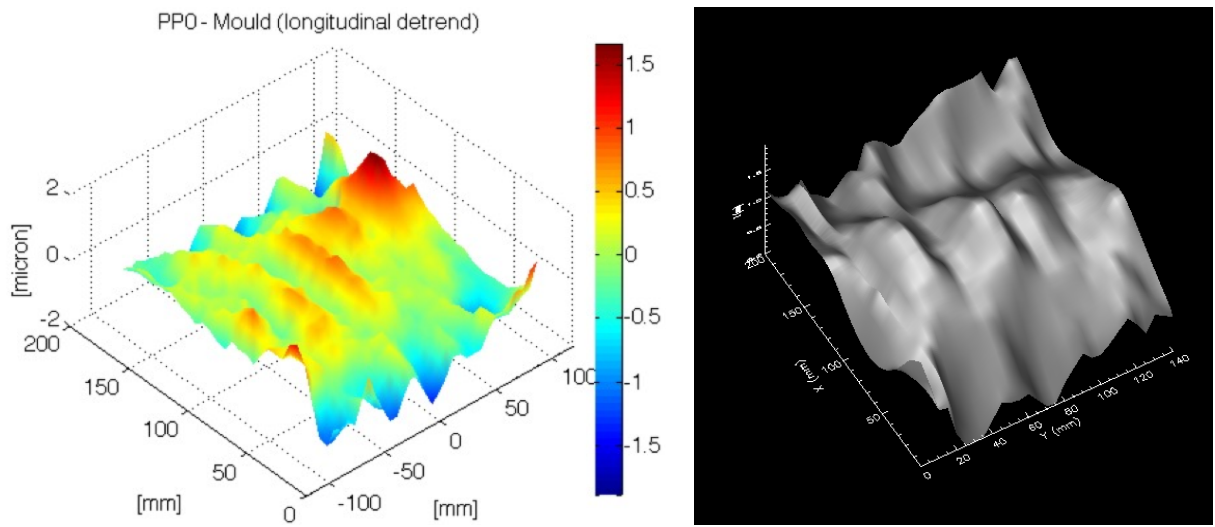

Fig. 2: (left) Error shape of the parabolic segment of the upper plate pair of the PoC#2, directly measured with the CUP (Characterization Universal Profilometer). The paraboloidal mould surface was subtracted[5]. (right) The surface map as reconstructed from the intrafocus X-ray (0.27 keV) image recorded at PANTER[22] (see Sect. 4).

We explicitly note that a relevant improvement was obtained throughout the demonstrators manufactured (see Tab. 1). The typical HEW range of the X-ray optics replicated by Nickel elctroforming was reached with hot slumped glasses, but improvements below 20 arcsec HEW will require a more accurate slumping figure. In parallel, a possibility to be envisaged is the *active correction* of the figure at the locations where the error is the largest, i.e., between the ribs. In Sect. 4 we will see that quantitative information on the mirror deformation map can be directly extracted from the analysis of the intra-focal image, and so offers the opportunity to feed the piezoelectric array with the voltage needed.

### 2.1. Thin plastic foil optics

The possibility to manufacture X-ray optics with a thin plastic foil have been studied in the last years[7] by a collaboration between SAO, the Palermo University, INAF/OAPA, and DSRI. The obvious advantage of plastic is that it is a lightweight material and very simply to shape. It has also proven to well withstand the deposition of a metallic coating and also of a multilayer. Finally, plastics with a good smoothness[10] and good thermo-mechanical properties can be selected[8],[9]. Possible geometries explored so far are the cylindrical, the conical and the spiral one. Conversely, plastic foils are very easy to deform, so they need to be supported by a stiff, but not over-constraining, integration structure (Fig. 3). A first integration approach of cylindrical mirrors has been tried (Fig. 3, left). The integration process makes use of a figured mandrel to form cylindrical or conical shells from coated foils. The plastic foil is hold against the mandrel surface by vacuum chuck: the foil is then held inside grooves of the spokes of a supporting wheel. Epoxy is cured while foils are still inside the mandrel. An evolution of this process does not make use of epoxy resin but simply constrains the foil mechanically with vertical pins defining the outer geometry (Fig. 3, right). The surface tightness of the foils can be tuned to improve the azimuthal profile.

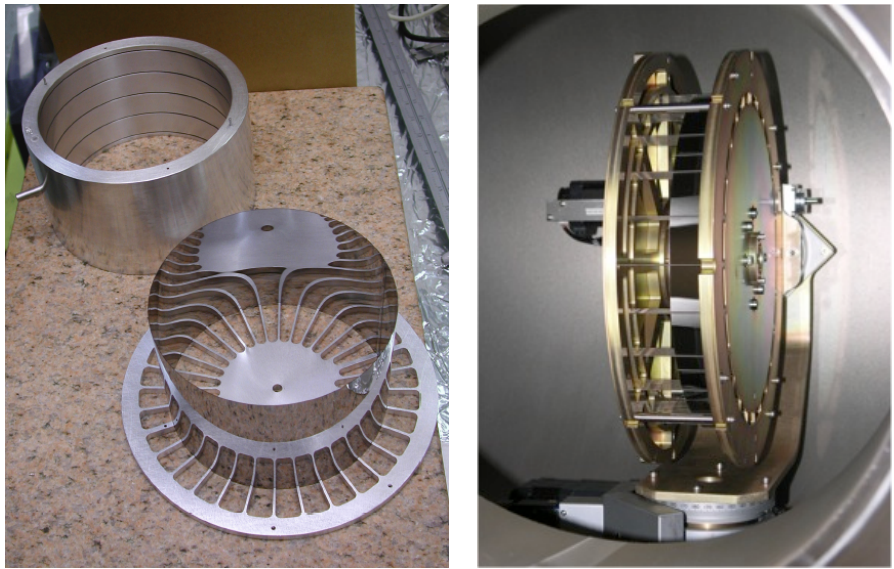

Fig. 3: (left) A thin plastic foil optic with tungsten coating mounted on the spokes of a spider structure. The mandrel used for integration is also visible in the background[8]. (right) A second integration concept, in which pins constrain the surface without using epoxy resin[9].

The angular resolution achieved in X-rays has been directly tested at the 35 m long XACT facility[17]. Following the first integration approach, prototypes of cylindrical mirror shells with 8 m focal lengths have reached a FWHM of 30 arcsec in focus (Fig. 4, left): the daisy shape of the focal spot denotes a major impact of roundness errors in the figure. A finite element modeling shows that a similar effect can be expected from thermal stress, especially near the spokes (Fig. 4, right). The second integration concept has returned better results in terms of FWHM at 0.27 keV (13 arcsec), even if the HEW is worse (1.9 arcmin) as a result of surface scattering and mid-frequency slope errors.

In addition to a further development of the integration concept to improve the shape of plastic foils, a viable possibility is the application of piezoelectric actuators to correct the slope defects where they have the maximum amplitude, as highlighted by the FEM analysis. Also in this case, the observation of the intra-focal image yields useful indications on the real shape of the mirror, which can be used to return a feedback to the voltage matrix.

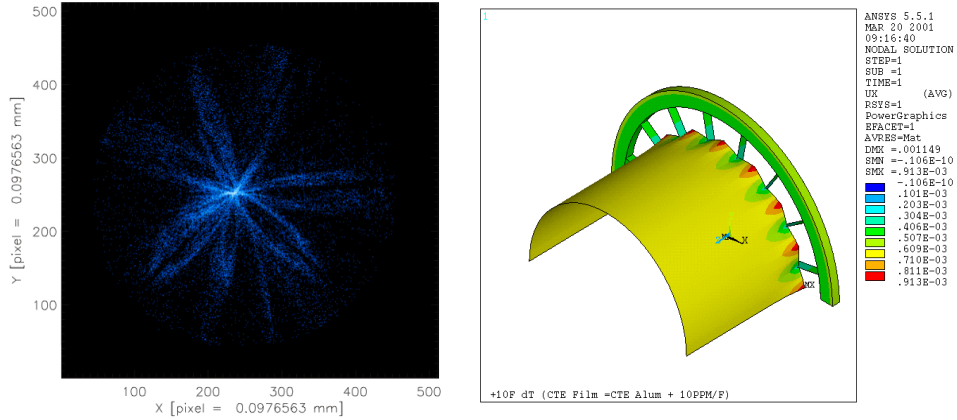

Fig. 4: (left) Measured focal spot[9] at the XACT facility, at 1.49 keV, of the mirror shell shown in Fig. 3, left. The "daisy" shape denotes the dominant presence of deformations in the azimuthal direction. (right) FEM analysis of the axial slope distortions expected from thermal stress in the model optic of Fig. 3, left[9].

## 3. MIRROR SHAPE CORRECTION VIA TANGENTIAL PIEZO ACTUATION

The AXYOM project (*Adjustable X-raY optics for astrOnoMy*) aims at improving via piezoelectric actuation the existing technologies for lightweight optics in glass or plastic foils, to a level of a few arcseconds HEW. In this phase, we are concentrating on the actuation of thin glass foil mirrors using commercial actuators, to be applied on the rear side of the glass foils at the integration stage, in the middle of two ribs, where the profile error has it maximum value. Unlike piezoelectric actuators commonly used in normal-incidence adaptive mirrors (which act in normal direction to their surface), the dense stacking of mirrors used in X-ray astronomical optics makes convenient to operate in a direction parallel to the surface. The piezo strain increases with the voltage applied and causes a local variation of the curvature radius, consequently correcting slope errors. The strain operates in both directions (longitudinal *and* azimuthal), but only the longitudinal direction needs to be corrected in general, while the unwanted correction in the sagittal direction has a negligible impact on the optical performances. Piezoelectric actuators of 3 types can be purchased:

1) *Piezoceramics transducers* (PZT) are commercially available (e.g. from *Physik Instrumente*) as bendable patches ready for use with built-in soldering pads, of variable sizes and thickness below 0.5 mm; hence of very small obstruction in a mirror stack. They are also cheap and exhibit high bending strength.
2) *Piezopolymers* (PVDF) are sold in thin (a few tens micron thick) plastic foils already poled and metallized (e.g., from *PiezoTech*). They can be easily trimmed to any size and moreover have already been used in space applications. Unfortunately, they are more expensive and the typical piezoelectric response for PVDF is 15 times lower than Piezoceramics.
3) *Microfiber Composites* (MFC) are also an interesting solution because are strongly anisotropic; i.e., they exhibit a high bending strength only in the longitudinal direction. They are, indeed, quite expensive.

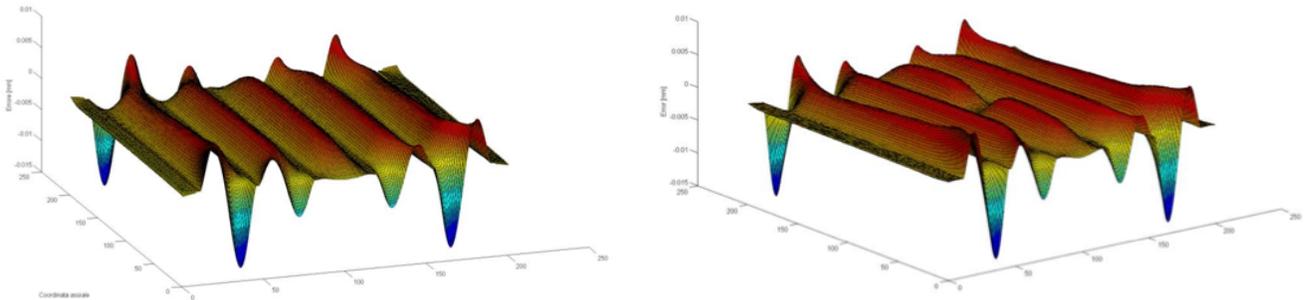

Fig. 5: (left) Finite Element Modeling of a cylindrical, 200 × 200 mm slumped glass with 1 m curvature radius and 0.4 mm thickness, simulated integration against a paraboloidal integration mould. Spring-back effect is clearly visible (as in Fig. 2) in the space between the ribs. (right) The same simulation with a single piezoceramic element P-876.SP1 by *PhysikInstrumente*, 16 mm × 16 mm wide and 0.2 mm thick, fed by a 10 V voltage: the springback effect was corrected locally. FEM performed via the Abaqus® software package (after[23]).

In order to operate a choice between the aforementioned actuators, we have implemented a Finite Element Model using a code based on Matlab® and Abaqus® for the integrated glass, initially assuming a perfectly slumped cylindrical foil. Hence, the modeling in absence of forces exerted on the backside of the glass simply includes the spring-back effect (Fig. 5, left). An actuator element with a 16 mm × 16 mm size has then been simulated in the center of the mirror; the computation has been performed assuming the piezoelectric constant values tabulated for different models of PZT or PVDF from the providers cited above, and the influence function on the mirror has been derived. The simulation is reported in detail in a previous SPIE volume[23]. Here we just mention the most important results:

i. The PVDF has a negligible influence on the glass foil.
ii. The influence of the PZT is inversely proportional to the thickness of the actuator.
iii. The PZT actuator model P-876.SP1 with a 200 μm thickness (Fig. 5, right) provides the strength necessary to locally correct the spring-back effect, even when energized at a moderate voltage (10 V).

These results suggest a first selection of the P-876.SP1 piezoelectric actuator as a good candidate for the development of the project. Other FEM simulations are currently in progress to simulate the effect on a real glass foil, characterized by a slumping error that is partly inherited by the integrated glass. Even if we can expect that the error to be corrected by the piezo array will be larger than the one simulated so far, this kind of PZTs can operate in a voltage from -100 to +400 V, so the dynamic range of available strains is more than one order of magnitude larger than in the example reported here.

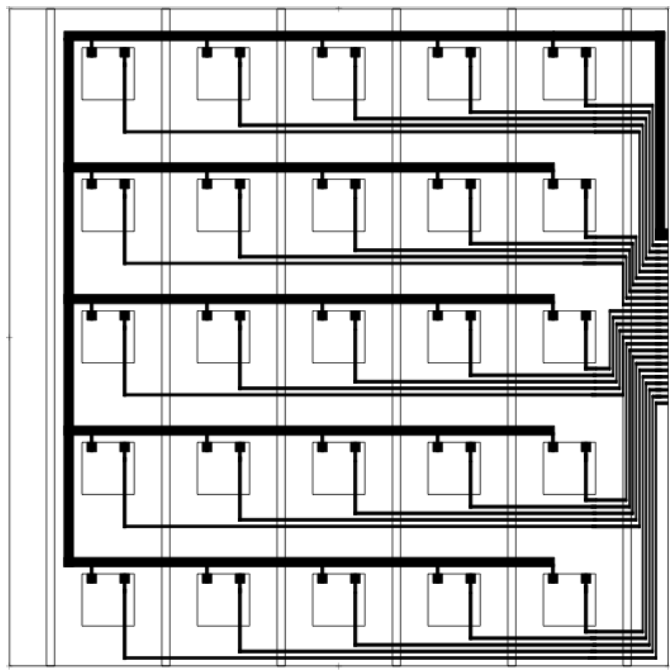

Fig. 6: A preliminary design of the electrical connections to be deposited on the backside of a slumped glass foil. The 6 stiffening ribs are shown in vertical. A 5 x 5 piezo matrix is assumed: the thick contact is grounded, the others are driven by a dedicated electronic. The glass outer sides, external to the ribs, are not actuated and will not be used as optical surface.

Another problem we are facing is how to bring the voltage signal to the PZT contacts without obstructing the space between consecutive shells in a XOU stack. Moreover, the contacts have to be extremely lightweight to avoid deformations of the glass. An option we are considering is the direct deposition of very thin contacts (in Au-Ti) on the non-optical side of the glass foil before the integration, using a lithographic process. A drawing of the electrical connections to be realized on a mirror prototype with a 5×5 array of P-876.SP1 actuators is shown in Fig. 6.

The last hardware issue is the electronic unit to be assembled to drive the voltage array. Active mirrors for reflecting telescopes traditionally make use of a considerable but limited number of high voltage piezo actuators, which are controlled by a multitude of single channel high voltage programmable generators. In our project, given the large number of actuators that will be placed on each mirror (25 to 50) and, in turn, the large number of active mirrors to be managed in a stack, an enormous number of control channels can easily build up. A different approach is thus being considered, where each active mirror plays the role of a "smart mirror". A dedicated electronic circuit will manage all the piezo actuators on a single mirror, making each mirror addressable as in a local network. The main advantage of this approach is a dramatic reduction in the number and length of wires, as well as a more organic management of data transfer. Two key points play in favor of this approach: the relatively relaxed constraints in term of control signal bandwidth, just a few Hertz; the moderate voltages needed for controlling the piezoceramics, which should be around few tens of volts[23].

The voltage signals that drive all the actuators for a single smart mirror will be multiplexed. Many challenges will have to be overcome during the project: the reduction of power consumption, which will be faced by using the latest generation of ultra low power integrated circuits; circuit interconnection issues, addressed by moving from traditional PCBs (Printed Circuit Boards) to miniaturized flexible Kapton PCBs; the choice between multiplexing the voltage signals which drive all the actuators for a single smart mirror versus the possibility to have a miniaturized voltage generator dedicated to each actuator.

## 4. FEEDBACK VIA INTRAFOCUS IMAGING

One of the crucial problems in active optics is represented by the feedback. Even if the influence of each actuator on the mirror shape can be characterized at the LTP or using another surface mapping tool, and even if there are algorithms to determine the optimal piezoelectric voltages[24] for a measured shape to be corrected, the convergence relies on a metrological tool that cannot be always used in a densely stacked XOU. Moreover, it cannot be done in situ, in the real thermal and structural configuration experienced in the vacuum X-ray chamber used for testing, or even when the mirror is operated in space.

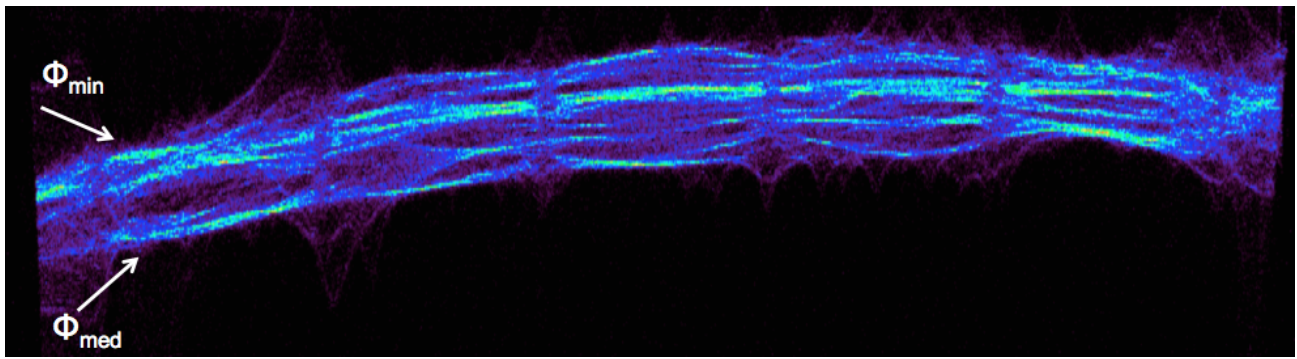

Fig. 7: Intra-focus (at 7.8 m from the mirror exit pupil) image of the hyperbolic segment of a layer of the PoC#2 (Fig. 1, right), recorded at the PANTER facility[22], in single reflection setup (after[15]). The 133 mm-long arc is a superposition of a 14-exposure scan with the TRoPIC detector. The striations along the azimuth are related to mid-frequency errors in the axial profiles. The 6 rib locations are clearly seen where the trace is narrower and the striations tend to disappear.

A solution comes from the direct observation of the focal spot in X-rays. In principle, one might try to minimize the HEW measured in X-rays changing the voltage combinations until a minimum is found. Owing to the large number of actuators at play, however, such a blind search is expectedly very complex. Near the focus, all the features coming from different parts of the mirror overlap each other, and cannot be disentangled easily. However, in intra-focus position, near enough to the exit pupil, the different parts of the mirrors become distinguishable (Fig. 7). The intra-focal trace is in general broader than the camera size; so, the entire arc has to be recorded by a camera scan.

The interesting aspect is the intra-focal trace of a real mirror is not uniform, but characterized by striations nearly parallel to the azimuthal direction. If we assume that the beam deviations out of the incidence plane are negligible (this is usually fulfilled with our glass mirrors), and if the mirror is close enough to avoid the rays to cross each other, it is possible to extract the profile error from the intensity variation of the trace in the radial direction (Fig. 8). The detailed method is exposed in a previous paper[15], where it has been applied to the intra-focus measurement (Tab.1) of the PoC#2 (Fig.1) at PANTER. Slicing the intra-focal trace of Fig. 7 and computing the intensity profile, the longitudinal profiles can be computed after a few iterations (Fig. 9). The longitudinal profiles are finally assembled into a surface error map. Provided that a mirror surface inside the stack can be seen in single reflection (e.g. by a proper XOU tilt), its shape can be characterized without a wavefront sensor. In this way, the voltage matrix can be computed and applied to the piezoelectric array to correct the measured figure error. A similar method can be used to solve beam-shaping problems in focus[25]. We note that a very similar approach[26] was in use in optical astronomy since the '80s, but – at least, to our knowledge - never applied to grazing incidence X-ray mirrors before the PoC#2 tests[15].

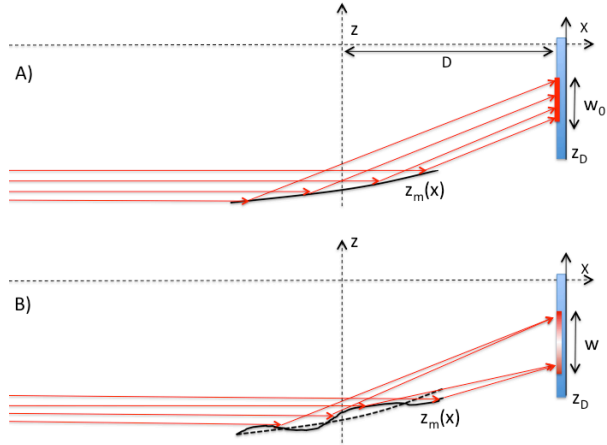

Fig. 8: The principle of the error profile reconstruction from the intensity observed on the detector in intra-focal position[15]. A) A perfect mirror would produce a perfectly uniform trace; B) Profile defects affect trace width and brightness variations.

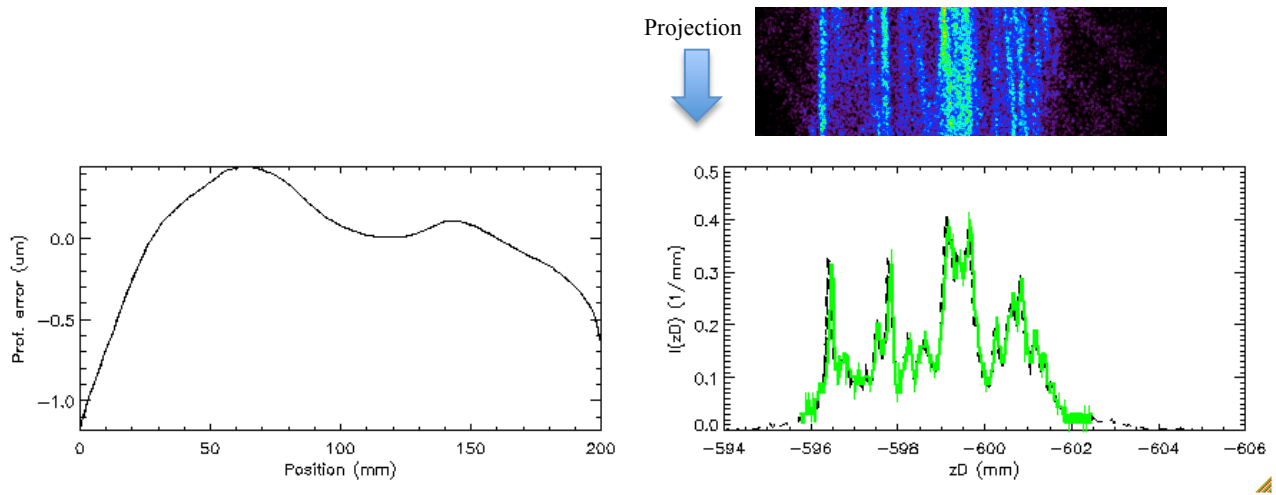

Fig. 9: (top) Central slice of the intra-focal trace shown in Fig. 7. (bottom right) The projected intensity (dashed line) superposed to the ray-tracing findings (color line) from the reconstructed profile (bottom left). After[15].

For this project, intra-focal X-ray tests of mirror demonstrators will be performed at the 35 m long X-ray facility XACT[17] at INAF/OAPA (Fig. 10). The XACT facility includes:

- Electron impact X-ray sources covering the range 0.1-20 keV;
- X-ray monochromators;
- a 35 m high vacuum pipe, including 3 test chambers;
- Vacuum Micropositioning systems:
- X-ray detectors (4 cm diam. Microchannel Plate);
- Vacuum Alt-azimuth mount.

Exactly like we did at PANTER with the PoC#2[15], the intra-focal pattern will be used to reconstruct the 3D map of the actuated mirror under X-rays. Since the influence function of each actuator has been modeled, or measured if accessible to usual metrology tools (but it is unlikely measured directly for a mirror inside a stack), we will be able to calculate[24] which actuators should be fed with an appropriate voltage, using the dedicated electronic unit. The process can be repeated for a refinement of the shape correction.

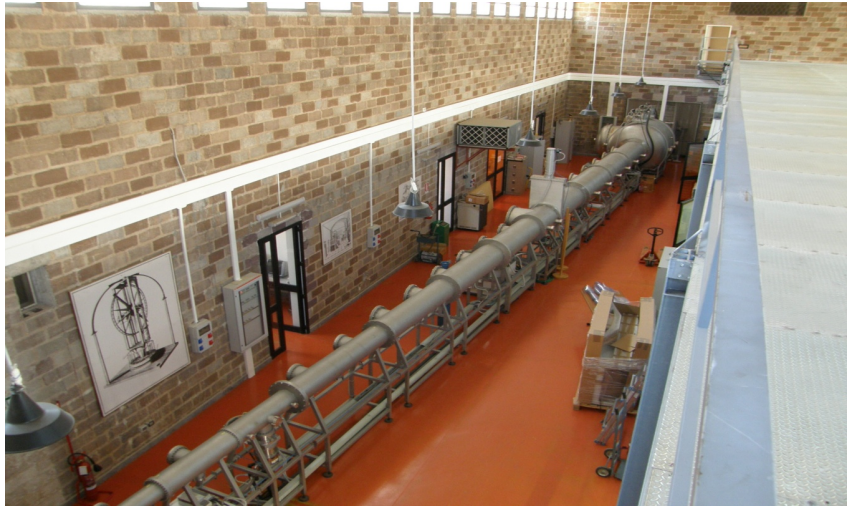

Fig. 10: View of the 35 m long XACT facility, at INAF/OAPA.

## 5. CONCLUSIONS

We are taking our first steps in active X-ray optics for astronomical applications. Our approach is the activation of lightweight X-ray optics already developed in INAF/OAB with slumped glasses, and in UNIPA on plastic foils, using commercial piezoelectric components, the metrology tools already available, and the existing test facility at INAF/OAPA. Finite Element Modeling has shown which model of piezoceramic component is suitable for the shape correction of integrated, slumped, 0.4 mm thick glass foils. The deposition of the electrical contacts on the rear side of the glass mirrors is also being realized. Feedback on the actual shape imparted by the piezoelectric array will be provided in X-ray full-illumination, in situ and real time, in intra-focal setup, at the XACT facility. FEM analysis is ongoing to extend the simulation to a mirror with a measured profile error, and to determine the optimal piezoceramic array geometry. Piezoceramic actuators will be purchased and characterized soon.

## ACKNOWLEDGMENTS

The AXYOM project, devoted to the study of the correction of thin glass/plastic foils for X-ray mirrors, is financed by a TECNO-INAF 2012 grant.